\documentclass[aps, pra, reprint, superscriptaddress, longbibliography]{revtex4-2}
\usepackage[utf8]{inputenc}
\usepackage[export]{adjustbox}
\usepackage{tikz}
\usepackage{amsmath}
\usepackage{amssymb}
\usepackage{hyperref}
\usepackage{graphicx}
\usepackage{epstopdf}
\usepackage{dcolumn}
\usepackage{soul}
\usepackage{mathrsfs}
\usepackage{bm}
\usepackage{soul}
\usepackage{xspace}
\usepackage{verbatim}
\usepackage{color}
\usepackage{xcolor}

\hypersetup{colorlinks=true,linkcolor=blue,citecolor=blue, filecolor=blue,urlcolor=blue,breaklinks=true}
\usepackage[T1]{fontenc}

\begin{document}

\title{Sequential measurements thermometry with quantum many-body probes}

\author{Yaoling Yang}
\email{yyaoling@std.uestc.edu.cn}
\affiliation{Institute of Fundamental and Frontier Sciences, University of Electronic Science and Technology of China, Chengdu 611731, China}

\author{Victor Montenegro}
\email{vmontenegro@uestc.edu.cn}
\affiliation{Institute of Fundamental and Frontier Sciences, University of Electronic Science and Technology of China, Chengdu 611731, China}
\affiliation{Key Laboratory of Quantum Physics and Photonic Quantum Information, Ministry of Education, University of Electronic Science and
Technology of China, Chengdu 611731, China}

\author{Abolfazl Bayat}
\email{abolfazl.bayat@uestc.edu.cn}
\affiliation{Institute of Fundamental and Frontier Sciences, University of Electronic Science and Technology of China, Chengdu 611731, China}
\affiliation{Key Laboratory of Quantum Physics and Photonic Quantum Information, Ministry of Education, University of Electronic Science and
Technology of China, Chengdu 611731, China}

\date{\today}

\begin{abstract}
Measuring the temperature of a quantum system is an essential task in almost all aspects of quantum technologies. 
Theoretically, an optimal strategy for thermometry requires measuring energy which demands full accessibility over the entire system as well as complex entangled measurement basis. In this paper, we take a different approach and show that single qubit sequential measurements in the computational basis not only allows precise thermometry of a many-body system but may also achieve precision beyond the theoretical bound, avoiding demanding energy measurements at equilibrium.
To obtain such precision, the time between the two subsequent measurements should be smaller than the thermalization time so that the probe never thermalizes. Therefore, the non-equilibrium dynamics of the system continuously imprint information about temperature in the state of the probe. This  allows the  sequential measurement scheme to reach precision beyond the accuracy achievable by complex energy measurements on equilibrium probes. 
\end{abstract}

\maketitle

\section{Introduction}

The estimation of temperature, known as thermometry~\cite{De_Pasquale_2018, Mehboudi_2019}, holds relevance across all branches of natural sciences~\cite{D2CS00069E, C2NR30663H, Kucsko2013, Fujiwara_2021, nano13212904,chu2022thermodynamic}. To date, its significance becomes increasingly apparent in numerous physical applications, particularly those demanding low-energy excitations within cryogenic environments. Indeed, bringing the system to ultracold temperatures allows us for observing quantum effects or executing precise processing tasks~\cite{Huang2020}. Hence, a precise thermometry scheme is crucial for determining, for instance, the impact of quantum features in the presence of thermal fluctuations~\cite{PhysRevA.108.012433, PhysRevResearch.6.013001}. So far, thermometry has been explored in several contexts such as non-linear optomechanics~\cite{PhysRevA.92.031802, PhysRevResearch.2.043338}, single particle systems~\cite{Razavian2019, Burgarth2015, oconnor2024fisher}, solid state impurity systems~\cite{mihailescu2023thermometry,mihailescu2023multiparameter}, topological  spinless fermions~\cite{srivastava2023topological}, 
establishing general bounds for optimal nonequilibrium thermometry~\cite{Sekatski_2022}, probe optimization~\cite{PhysRevLett.114.220405, Mukherjee_2019, glatthard2022optimal}, global thermometry~\cite{mok2021optimal, rubio2021global}, in coupled harmonic oscillators~\cite{campbell2017global}, collisional models~\cite{alves2022bayesian}, thermodynamic length~\cite{jorgensen2022bayesian}, and ultimate thermometry bounds for arbitrary interactions and measurement schemes~\cite{mehboudi2022fundamental}.

Conventional thermometry relies on the zeroth law of thermodynamics~\cite{Razavian2019} which implies that a probe reaches an equilibrium Gibbs state $\rho(T) = \exp\{-H/k_BT\}/\mathcal{Z}$, 
due to interaction with a thermal reservoir at temperature $T$. Here, $H$
is the Hamiltonian of the probe, $k_B$ is the Boltzmann constant and $\mathcal{Z}$ is the partition function. 
For any given probe, the best temperature estimation can be obtained by energy measurement, namely $H$ being the observable to be measured~\cite{PhysRevE.83.011109, PhysRevA.82.011611}. Such measurement results in  ultimate precision of thermometry, quantified by the variance of temperature $\mathrm{Var}[T]$, which is given by  $\mathrm{Var}[T] \geq T^{4}/M(\Delta H)^2$, where $M$ is the number of measurements and $(\Delta H)^2$ is the energy fluctuations~\cite{PhysRevE.83.011109, Mehboudi_2019, De_Pasquale_2018}. 
In this context, studies on optimal thermometry strategies with coarse-grained measurements have been pursued~\cite{Hovhannisyan_2021, oconnor2024fisher} and even beyond standard open-system weak-coupling assumptions~\cite{Glatthard_2023}. One may focus on finding the optimal probe by engineering the Hamiltonian $H$. Such optimization for a local estimation scheme, where prior information about the temperature is available,  results in an effective two-level system with a maximally degenerate excited state~\cite{PhysRevLett.114.220405}. In the absence of prior information, i.e. global sensing, the optimal probe becomes more complex and extra energy levels are also required~\cite{mok2021optimal}. 
The fact that conventional optimal thermometry requires energy measurement makes it very challenging as such actions require: (i) full accessibility over the entire system; and (ii) complex entangled measurement basis. To overcome these challenging requirements, we investigate a non-conventional metrology scheme applied to thermometry by exploiting sequences of local quantum measurements on the probe at consecutive time intervals~\cite{PhysRevLett.129.120503, Burgarth2015, PhysRevResearch.5.043273}.

Quantum measurement and its subsequent wave function collapse can be used as means for inducing non-equilibrium dynamics in a many-body system at equilibrium~\cite{Busch1990, Schmidt_2020, BAN2021127383, PhysRevX.9.031009, PhysRevLett.128.010604, benoist2023limit, Haapasalo_2016, benoist2019invariant, burgarth2014exponential, pouyandeh2014measurement, pouyandeh2014measurement, Bayat, ma2018phase, PhysRevA.92.042315}. Consecutive local measurements, each separated by a period of non-equilibrium dynamics, provide information about the underlying many-body system which can be used for sensing purposes~\cite{HMabuchi_1996,Burgarth2015, PhysRevA.96.012316, Ritboon_2022, bompais2023asymptotic, 9992617, proceedings2019012011, PhysRevA.64.042105, PhysRevA.94.042322, Nagali_2012, PhysRevA.92.032124,PhysRevLett.129.120503,PhysRevResearch.5.043273, PhysRevA.99.022102, Radaelli_2023}. In thermometry context, several open questions may arise. Can one achieve precise thermometry through local measurements? If so, can such measurements surpass the precision of conventional optimal thermometry, through exploiting non-equilibrium dynamics? Indeed, a natural question is whether by repeating enough measurements one can reach thermometry precision beyond the complex energy measurement in equilibrium probes.

In this work, we estimate the temperature of a  many-body probe which is described by  Heisenberg interaction. A sequence of consecutive measurements, each followed by a period of non-equilibrium dynamics, is used for estimating the temperature of the system. Three different regimes have been studied, namely weak, strong and  intermediate thermalization rates. In the intermediate regime where the thermalization rate is comparable with the exchange coupling between the particles our sequential measurement strategy 
surpasses the conventional optimal thermometry approach, where the whole thermalized probe is measured in energy basis.

The rest of the paper is organized as follows: in Sec.~\ref{sec_standard_bounds}, we introduce the figures of merit for quantum parameter estimation. In Sec.~\ref{sec_thermo_process}, we introduce the model of the quantum many-body probe and its open dynamics between the probe and the bath. In Sec.~\ref{sec_seq_measurements}, we outline the procedure for sequential measurements sensing. In Sec.~\ref{sec_thermo_regimes}, we study the weak, intermediate, and strong thermalization regimes. Finally, in Sec.~\ref{sec_conc}, we conclude our work.

\section{Information Metrics}\label{sec_standard_bounds}
The uncertainty in estimating an unknown parameter $\lambda$ encoded into a quantum state $\rho(\lambda)$ satisfies the Cram\'{e}r-Rao inequality~\cite{Holevo, cramer1999mathematical, LeCam-1986, helstrom1969quantum,zamir1998proof}
\begin{equation}
    \mathrm{Var}[\lambda]\geq \frac{1}{M\mathcal{F}},\label{eq_cramerrao_classical}
\end{equation}
where $\mathrm{Var}[\lambda]$ is the variance of $\lambda$, $M$ denotes the number of measurement trials, and $\mathcal{F}$ stands for the classical Fisher information~\cite{nla.cat-vn81100, Rao1992}:
\begin{equation}
    \mathcal{F}= \sum_j \frac{1}{p_j(\lambda)} \left[\partial_\lambda p_j(\lambda)\right]^2.\label{eq_cfi}
\end{equation}
Here, $\partial_\lambda:=\frac{\partial}{\partial\lambda}$ and the summation $\sum_j$ runs over all $j$ countable measurement outcomes with associate probability $p_j(\lambda)=\mathrm{Tr}[\Pi_j\rho(\lambda)]$, where $\Pi_j$ is a positive operator-valued measure (POVM) with random outcome $j$. The classical Fisher information has a well-defined operational meaning, namely: for any fixed measurement basis, the achievable uncertainty is at best lower bounded by the right-hand side of the Cram\'{e}r-Rao inequality in Eq.~\eqref{eq_cramerrao_classical}, with the best scenario achievable if an optimal estimator is employed.

In single-parameter sensing, the optimal measurement that maximizes the classical Fisher information is termed the quantum Fisher information, denoted as $\mathcal{Q} = \max_{\{\Pi_j\}}[\mathcal{F}]$~\cite{paris2009quantum}. Hence, by definition, the Cram\'{e}r-Rao inequality updates to~\cite{HELSTROM1967101, paris2009quantum, 1055103, 1055173, PhysRevLett.72.3439}:
\begin{equation}
    \mathrm{Var}[\lambda]\geq \frac{1}{M\mathcal{F}}\geq \frac{1}{M\mathcal{Q}}.\label{eq_cramerrao_quantum}
\end{equation}
The quantum Fisher information $\mathcal{Q}$, as an optimization procedure over all possible POVMs, can also be formulated in terms of the Symmetric Logarithmic Derivative (SLD) self-adjoint operator $\mathcal{L}(\lambda)$~\cite{paris2009quantum}. This SLD operator is a solution of the Lyapunov equation $2\partial_\lambda\rho(\lambda) = \{\mathcal{L}(\lambda), \rho(\lambda)\}$, where $\{\cdot, \cdot\}$ denotes the anticommutator. In general, it can be proven that the quantum Fisher information can be computed as~\cite{paris2009quantum}:
\begin{equation}
    \mathcal{Q} = \mathrm{Tr}[\rho(\lambda)\mathcal{L}(\lambda)^2]= \mathrm{Tr}[\partial_\lambda\rho(\lambda)\mathcal{L}(\lambda)].\label{eq_qfi}
\end{equation}
The quantum Fisher information has also a well-defined operational, namely: it represents the ultimate sensing precision achievable to estimate the unknown parameter $\lambda$ encoded in the quantum probe $\rho(\lambda)$. Note that to achieve the ultimate level of precision, one must calculate the SLD $\mathcal{L}(\lambda)$ in Eq.~\eqref{eq_qfi}. However, $\mathcal{L}(\lambda)$ depends on the unknown parameter, and consequently, the SLD poses significant challenges in local estimation theory, requiring substantial prior information about the unknown parameter~\cite{Hajek1970, LeCam-1986}. This constraint has been addressed in global sensing scenarios~\cite{vantrees1968, bj/1186078362, Montenegro2021}, eliminating the need for prior information about the parameter. Furthermore, even if the optimal measurement is identified, through the eigenstates of the SLD~\cite{paris2009quantum}, it can be particularly demanding in practice. Hence, a more practical direction can be pursued by exploring sensing with a readily accessible measurement basis.

\section{Thermalization Process}\label{sec_thermo_process}
We consider a one-dimensional many-body probe of $N$ interacting spin-1/2 particles with Heisenberg interaction. This is a paradigmatic model in magnetism, explored in both ground state~\cite{Mikeska2004} and non-equilibrium dynamics~\cite{PhysRevA.81.012304, PhysRevLett.91.207901, DeChiara_2006, PhysRevLett.32.170}. The Hamiltonian, with open boundary conditions, is:
\begin{equation}
    H = -J \sum_{j=1}^{N-1}\bm{\sigma}^j\cdot\bm{\sigma}^{j+1}, \label{eq_heisenberg_hamiltonian}
\end{equation}
where $J$ is the exchange interaction between particles and $\bm{\sigma}^j$ is a vector composed of Pauli matrices $\bm{\sigma}^j = (\sigma_x^j, \sigma_y^j, \sigma_z^j)$ at site $j$. Note that spin components obey $[\sigma^n_\alpha,\sigma^m_\beta]=2i\delta_{nm}\epsilon_{\alpha\beta\theta}\sigma^j_\theta$ where $(\alpha, \beta, \theta = x, y, z)$, $\delta_{nm}$ is the Kronecker delta, and $\epsilon_{\alpha\beta\theta}$ is the Levi-Civita symbol. Without loss of generality, we consider the ferromagnetic scenario throughout our work, wherein $J>0$. In this scenario, the energy is minimized when the spins are parallel to each other, being the triplet state favoured.

To simulate the open dynamics between the quantum many-body probe and the bath, we consider a microscopic master equation in Born-Markov approximation as its steady-state guarantees the true probe's thermalization~\cite{PhysRevA.75.013811, rivas2012open, Cattaneo_2019, Eremeev_2012, Montenegro_2011, Montenegro_2012}, given by:
\begin{multline}
    \dot{\rho}(t)=-i[H, \rho(t)] \\
    {+}\sum_{\substack{j{=}1\\ \omega{>}0}}^N\kappa(\omega)[A_j(\omega)\rho(t)A_j^\dagger(\omega){-}\frac{1}{2}\{A_j^\dagger(\omega)A_j(\omega),\rho(t)\}] \\
    {+}\sum_{\substack{j{=}1\\ \omega{>}0}}^N\kappa(-\omega)[A_j^\dagger(\omega)\rho(t)A_j(\omega){-}\frac{1}{2}\{A_j(\omega)A_j^\dagger(\omega),\rho(t)\}],\label{microscopic-master}
\end{multline}
where $H$ is the probe's Hamiltonian of Eq.~\eqref{eq_heisenberg_hamiltonian}. Note that, in Eq.~\eqref{microscopic-master}, each qubit is individually coupled to a common bath. To derive the quantum jump operators $A(\omega)$ for each spin $j$, we assume that each spin interacts with an infinite number of modes $\{b_k\}_{k=1,\ldots,\infty}$ with Hamiltonian
\begin{equation}
H_\mathrm{int}=\sum_{j=1}^N\sum_{k=1}^\infty g_k^j\sigma_x^j(b_k^\dagger + b_k), \label{eq_int_hamiltonian}
\end{equation}
with $g_k^{(j)}$ is the coupling strength between a local $j$th spin with the bath. Hence, the quantum jump operators $A_j(\omega)$ can be written in the form~\cite{PhysRevA.75.013811, rivas2012open, Cattaneo_2019, Eremeev_2012, Montenegro_2011, Montenegro_2012}

\begin{equation}
    A_j(\omega=\epsilon_m - \epsilon_n)=|\epsilon_n\rangle\langle\epsilon_n|\sigma_x^j|\epsilon_m\rangle\langle\epsilon_m|,
\end{equation}
where $H|\epsilon_l\rangle=\epsilon_l|\epsilon_l\rangle$ ($l=0,1,\ldots,2^N$) and $\omega= \epsilon_m - \epsilon_n > 0$ is the Bohr transition frequency from energy level $\epsilon_n$ to $\epsilon_m$.
Most notably, to ensure that the steady-state of the master equation, in the presence of energy losses, corresponds to the Gibbs state as described in Eq.~\eqref{eq_gibbs_state}, one imposes the Kubo-Martin-Schwinger (KMS) condition \cite{PhysRevA.75.013811, rivas2012open}:
\begin{equation}
    \kappa(-\omega) = \mathrm{exp}\left(-\frac{\omega}{T}\right)\kappa(\omega).
\end{equation}
Note that $\kappa(\omega)$ is generally computed as a function of the correlation function of the bath~\cite{rivas2012open}. Here, however, we assume that $\kappa(\omega)$ is a free parameter defined by a constant value $\kappa$ and thus $\kappa(-\omega) = \mathrm{exp}\left(-\frac{\omega}{T}\right)\kappa $.

\begin{figure}[t]
    \centering
    \includegraphics[width=\linewidth]{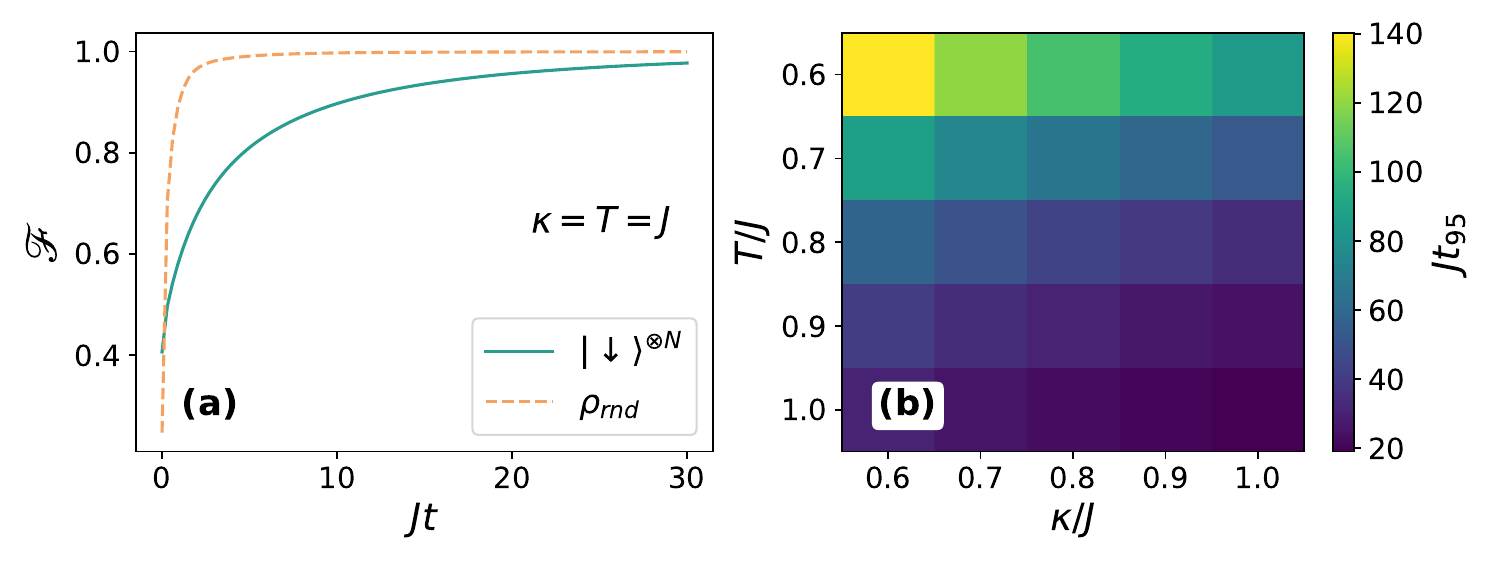}
    \caption{(a) The fidelity $\mathscr{F}$ between the Gibbs state $\rho_{\text{th}}$ and the state undergoing the thermalization process as a function of time. We consider $T=\kappa=J$. (b) The time $t_{95}$ required to achieve a fidelity greater than 95\% between the evolved state and the Gibbs state as functions of $\gamma$ and $T$, starting from the initial state $|\psi\rangle = |\downarrow\rangle^{\otimes N}$. The system size is fixed at $N=4$ for both figures.}
    \label{fig_thermalization}
\end{figure}

To demonstrate that the master equation in Eq.~\eqref{microscopic-master} effectively thermalizes the system into the correct Gibbs state, we evaluate the fidelity $\mathscr{F}$ between the Gibbs state $\rho_{\text{th}}=e^{-H/k_BT}/\mathcal{Z}$ and a state evolving under the dynamics governed by Eq.~\eqref{microscopic-master}. In Fig.~\ref{fig_thermalization}(a), we plot the fidelity $\mathscr{F}$ between $\rho_{\text{th}}$ and $|\psi\rangle = |\downarrow\rangle^{\otimes N}$ and a randomly generated initial state $\rho_{\text{rnd}}$ as a function of time. Without loss of generality, we consider $T=\kappa=J$, and $J=1$. As the figure shows, the fidelity approaches unity for both initial states; therefore, the open dynamics thermalizes into the correct Gibbs state of the system. In Fig.~\ref{fig_thermalization}(b), we plot the time $t_{95}$ needed to reach a fidelity $\mathscr{F}\geq 95$ as functions of $\kappa$ and $T$. Without loss of generality, we consider the fidelity between $\rho_{\text{th}}$ and the initial state $|\psi\rangle = |\downarrow\rangle^{\otimes N}$ for $N=4$. As seen from the figure, both higher temperatures $T$ and larger thermalization rates $\kappa$ thermalizes the probe faster.

With these settings, we identify three regimes of thermalization, namely: (i) weak thermalization regime $\kappa\ll J$; (ii) intermediate thermalization regime $\kappa \sim J$; and (iii) strong thermalization regime $\kappa \gg J$. Each case will be address using sequential measurements thermometry in the following sections.

\section{Sequential Measurement Sensing Protocol}\label{sec_seq_measurements}
\begin{figure}[t]
    \centering
    \includegraphics[width=\linewidth]{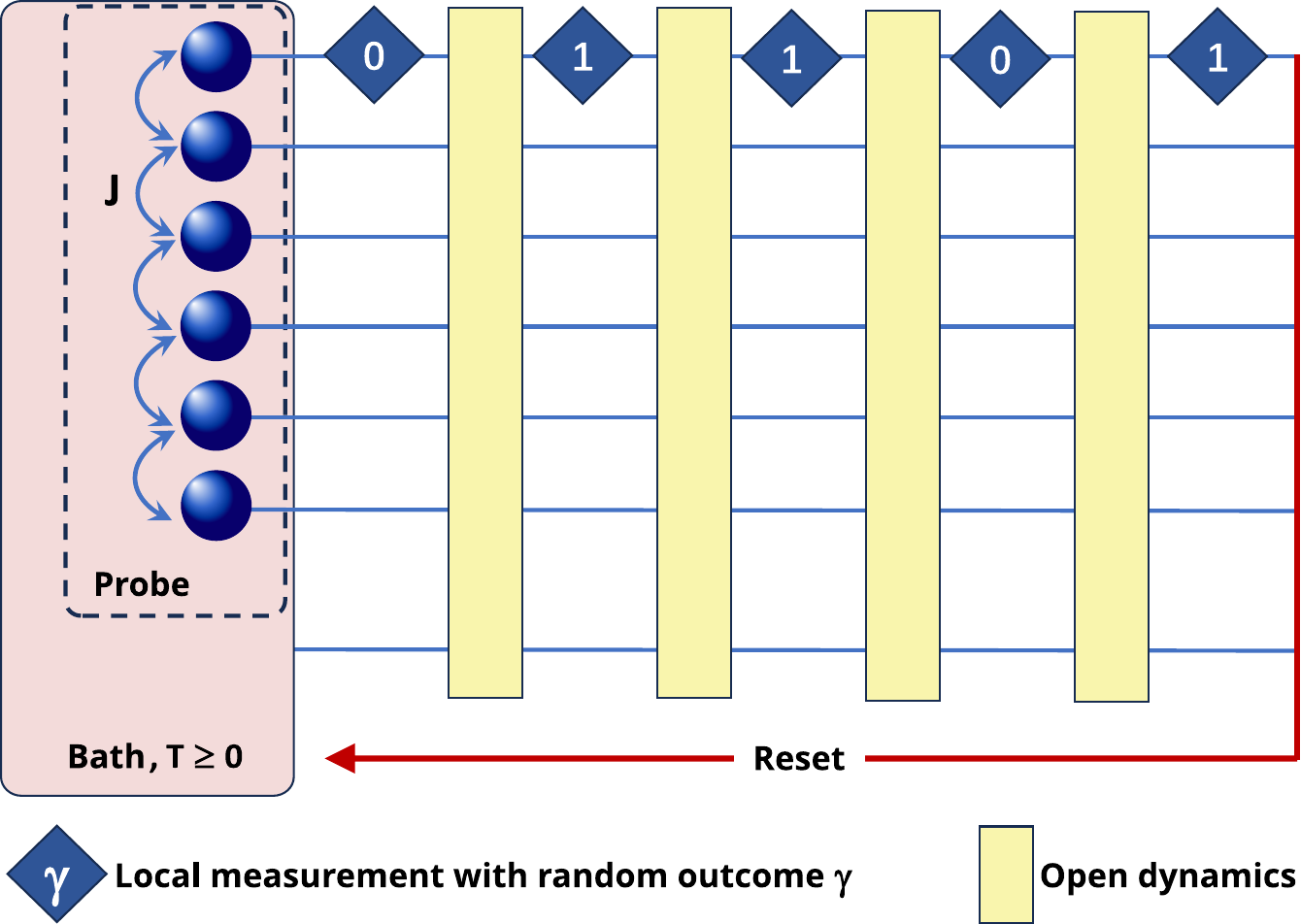}
    \caption{Sketch of sequential measurements metrology adapted for thermometry purposes: A quantum many-body probe with exchange interaction $J$ is initialized at thermal equilibrium with a bath at an unknown temperature $T$. A finite number of local sequential measurements $n_\mathrm{seq}$ are performed on the probe, followed by the probe-bath open evolution, collecting correlated measurement outcomes. Here, the single quantum trajectory $\pmb{\gamma}_1$ collects five measurement outcomes $\pmb{\gamma}_1=(0, 1, 1, 0, 1)$ corresponding to $n_\mathrm{seq}=5$. After a certain number of sequential measurements are performed on the probe, the system is reset for a new quantum trajectory. }
    \label{fig_schematics}
\end{figure}

The standard approach to estimating an unknown parameter, such as temperature, involves conducting a single measurement on each identical copies of the probe to construct probability distributions. Once the probability distributions are determined, once can evaluate the classical Fisher information as shown in Eq.~\eqref{eq_cfi}. Here, we adopt an unconventional sensing approach, where a sequence of measurements are performed on the probe before restarting the system for a new run of sequential measurements. This sensing protocol collects correlated measurement outcomes, which reduces the number of identical copies of the probe (or the total protocol time) to achieve the same level of sensing precision. Thus, using the sensing resources significantly more efficiently~\cite{PhysRevLett.129.120503, PhysRevResearch.5.043273}. Upon conducting a local measurement on the probe, the  quantum state of the entire system collapses, effectively creating another probe that evolves during a successive time interval. The steps of the sequential measurements sensing procedure are as follows, see Fig.~\ref{fig_schematics}:
\begin{enumerate}
    \item[(i)] A quantum probe evolves according to Eq.~\eqref{microscopic-master} from $\rho^{(j)}(0)$ to $\rho^{(j)}(\tau_j)$,
    
    \item[(ii)] At time $\tau_j$ a POVM $\{\Pi_{\gamma_j}\}$ with random outcome $\gamma_j$ is performed on the probe, collapsing the state into 
    \begin{equation}
        \rho^{(j{+}1)}(0){=}\frac{\Pi_{\gamma_j}\rho^{(j)}(\tau_j)\Pi_{\gamma_j}^\dagger}{p(\gamma_j)},
    \end{equation}
    where 
    \begin{equation}p(\gamma_j){=}\text{Tr}[\Pi_{\gamma_j}\rho^{(j)}(\tau_{j})\Pi_{\gamma_j}^\dagger],
    \end{equation}
    is the probability associated to $\gamma_j$ at step $j$,
    
    \item[(iii)] The random outcome $\gamma_j$ is recorded and the new initial state $\rho^{(j{+}1)}(0)$ is replaced in (i),
    
    \item[(iv)] The above steps are repeated until $n_\mathrm{seq}$ measurements outcomes are consecutively obtained,
    
    \item[(v)] After gathering a data sequence of length $n_\mathrm{seq}$, i.e., $\pmb{\gamma}^{(n_\mathrm{seq})}{=}(\gamma_1,{\cdots},\gamma_{n_\mathrm{seq}})$, the probe is reset to $\rho_0^{(j)}$ and the process is repeated to generate a new trajectory.
\end{enumerate}

The above steps show that one effectively deals with different probes at each measurement step. Unlike standard sensing schemes~\cite{Degen, Giovannetti2011, Giovannetti2006, Giovannetti2004}, there is no requirement for specific maximally entangled probes, the necessity for exploiting quantum phase transitions, or quantum control. One recovers the standard sensing scheme by setting $n_\mathrm{seq} = 1$.

In the context of the sequential measurements sensing scenario, the relevant figure of merit is the classical Fisher information that accounts for the total information content provided by all possible quantum trajectories \cite{PhysRevResearch.5.043273}, namely:
\begin{equation}
\mathcal{F}^{(n+1)}=\mathcal{F}^{(n)}{+}{\Delta}\mathcal{F}^{(n+1)},\label{eq_CFI_trajectories}
\end{equation}
where $\mathcal{F}^{(n)}$ is the classical Fisher information at measurement step $n$, and ${\Delta}\mathcal{F}^{(n+1)}$ is the classical Fisher information increment after an additional measurement has been performed on the probe, that is:
\begin{equation}
\Delta\mathcal{F}^{(n+1)}:=\sum_{\pmb{\gamma}^{(n)}} P_{\pmb{\gamma}^{(n)}} f^{\pmb{\gamma}^{(n)}}.
\end{equation}
Here, $\sum_{\pmb{\gamma}^{(n)}}$ runs over all possible quantum trajectories of length $n$ and
\begin{equation}
    P_{\pmb{\gamma}^{(n)}}=\prod_{j=1}^n p(\gamma_j)
\end{equation}
is the conditional probability associated with a particular quantum trajectory of length $n$, i.e., $\pmb{\gamma}^{(n)}$. Note that while Eq.~\eqref{eq_CFI_trajectories} is the exact classical Fisher information for the sequential measurements sensing case, the exponential growth in computing all probability distributions limits the exact evaluation to $n_\mathrm{seq}\sim 20$. To explore the sensing capabilities of the probe for large values of $n_\mathrm{seq}$, we employ the Monte Carlo methodology developed in Ref.~\cite{PhysRevResearch.5.043273} to evaluate Eq.~\eqref{eq_CFI_trajectories}, where the most likely probability distributions are naturally selected. This approximation results in:
\begin{equation}
\Delta \mathcal{F}^{(n)} \sim \frac{1}{\mu}\sum_{j{=}1}^{\mu} \mathcal{F}^{j, (n)}, \label{eq_mc_delta_f}
\end{equation}
where $\mathcal{F}^{j,{(n)}}$ represents the classical Fisher information obtained from $p(\gamma_n)$ in Monte Carlo trajectory $j$ at measurement step $n$, and $\mu$ accounts for the total number of Monte Carlo samples.

\section{Sequential Measurements Thermometry}\label{sec_thermo_regimes}
The probe's initialization, i.e. before any measurement is performed on the probe, is adequately described by the Gibbs state
\begin{equation}
    \rho(T) = \frac{e^{-H/k_BT}}{\mathcal{Z}(T)}, \label{eq_gibbs_state}
\end{equation}
where $k_B$ denotes the Boltzmann constant (hereafter taken as $k_B=1$), $T$ is the temperature to be estimated, and $\mathcal{Z}(T)=\mathrm{Tr}[e^{-H/T}]$ is the partition function.

Note that it has been shown that the SLD for the probe at thermal equilibrium in Eq.~\eqref{eq_gibbs_state} is $\mathcal{L} = T^{-2}(H - \langle H \rangle)$~\cite{Mehboudi_2019}. Therefore, the quantum Fisher information reduces to:
\begin{equation}
    \mathcal{Q}=\frac{1}{T^{4}}(\Delta H)^2, \label{eq_qfi_heat}
\end{equation}
where $(\Delta H)^2 = \langle H^2 \rangle - \langle H \rangle^2$ is the energy fluctuations. This relates directly to the heat capacity of the system, denoted as
\begin{equation}
C_T := \partial_T \langle H \rangle = \frac{1}{T^{2}} (\Delta H)^2. \label{eq_heat_capacity}    
\end{equation}
Consequently, it can also be demonstrated that the most informative measurement regarding the temperature $T$ in Eq.~\eqref{eq_gibbs_state} are measurements in energy basis which require full accessibility to the entire system~\cite{PhysRevE.83.011109}. However, from a practical standpoint, having access to the whole system and energy measurements is not readily available. Therefore, determining the thermometry capabilities of a more minimalist approach, in which only local measurements are performed on a single qubit in the computational basis, is highly desirable. In what follows, we have all the ingredients to explore three relevant regimes of thermalization between the probe and the bath using a sequential measurements thermometry approach.

\subsection{Strong thermalization regime, $\kappa \gg J$} \label{subsec_thermal_equi}

\begin{figure}[t]
    \centering
    \includegraphics[width=\linewidth]{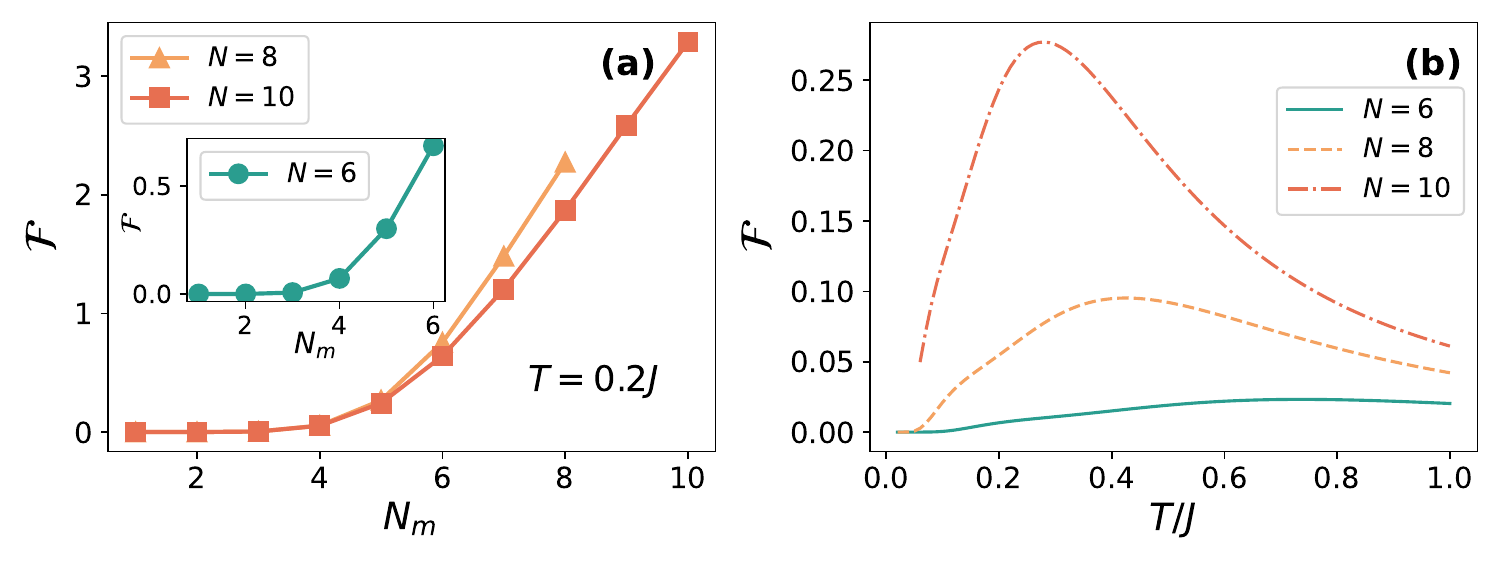}
\caption{(a) The classical Fisher information $\mathcal{F}$ as a function of the number of measured qubits $N_m$ for various system sizes $N$ for a fixed temperature $T=0.2J$. (b) The classical Fisher information $\mathcal{F}$ as a function of temperature $T$ when measuring half of the system for various system sizes $N$.}
    \label{fig_partial_measurements}
\end{figure}
In this scenario, the probe is strongly thermalized in between consecutive measurements. Thus, an approximation of the above is to consider the quantum many-body probe at thermal equilibrium---this case reduces to the particular case of sequential measurements with $n_\mathrm{seq}=1$, as a subsequent measurement will be performed on an identical copy of the probe due to its strong thermalization with the bath.

In Fig.~\ref{fig_partial_measurements}(a), we plot the classical Fisher information $\mathcal{F}$ as a function of the number of measured qubits $N_m$ at fixed temperature $T=0.2J$. The measurements are performed in the computational basis. As shown in the figure, the Fisher information $\mathcal{F}$ increases non-linearly with an increasing number of measured qubits. It is worth noting that local measurement on a single qubit $N_m=1$ gives zero Fisher information. Therefore, there exists a necessity to consider $N_m>1$ for estimation purposes. In Fig.~\ref{fig_partial_measurements}(b), we plot the classical Fisher information $\mathcal{F}$ as a function of temperature $T$ when measuring half of the system in the computational basis. As the figure shows, the Fisher information tends to vanish in two extreme cases: $T\rightarrow 0$ and $T\rightarrow\infty$, regardless of the system size. The value remains significantly smaller compared to the scenario of measuring more particles as shown in Fig.~\ref{fig_partial_measurements}(a).

To determine the ultimate sensing precision of the probe, we evaluate the quantum Fisher information as [see Eqs.~\eqref{eq_qfi_heat}-\eqref{eq_heat_capacity}]:
\begin{equation}
    \mathcal{Q} = \frac{1}{T^{2}}C_T = \frac{1}{T^{2}}\partial_T \langle H \rangle,
\end{equation}
where we have used the heat capacity shown in Eq.~\eqref{eq_heat_capacity}. While obtaining general analytical results for $\langle H \rangle$ and $\mathcal{Q}$ for the Heisenberg spin chain at thermal equilibrium poses challenges, it is possible to address the trivial case of $N=2$ and the first non-trivial scenario of $N=3$ ($N=4$ is also possible to obtain an analytical result, yet it is excessively cumbersome to provide any useful insight):
\[ 
\langle H \rangle = 
\begin{cases} 
    \frac{4J}{3 e^{4J/T}+1}-J & \text{for } N = 2 \\
    \frac{4 J \left(1-e^{\frac{6 J}{T}}\right)}{e^{\frac{4 J}{T}}+2 e^{\frac{6 J}{T}}+1} & \text{for } N=3,
\end{cases} 
\]
and
\[ 
\mathcal{Q} = 
\begin{cases} 
    12J^2 T^{-4} \left[\sinh \left(\frac{2J}{T}\right)+2 \cosh \left(\frac{2J}{T}\right)\right]^{-2} & \text{for } N = 2 \\
    \frac{8 J^2 e^{\frac{4 J}{T}} \left(9 e^{\frac{2 J}{T}}+e^{\frac{6 J}{T}}+2\right)}{T^4 \left(e^{\frac{4 J}{T}}+2 e^{\frac{6 J}{T}}+1\right)^2} & \text{for } N=3.
\end{cases} 
\]
Given that the behavior of $\mathcal{Q}$ remains universal regardless of the choice of $N$, the expressions presented for $N=2$ and $N=3$ suffice to provide significant insight.

\begin{figure}[t]
\includegraphics[width=\linewidth]{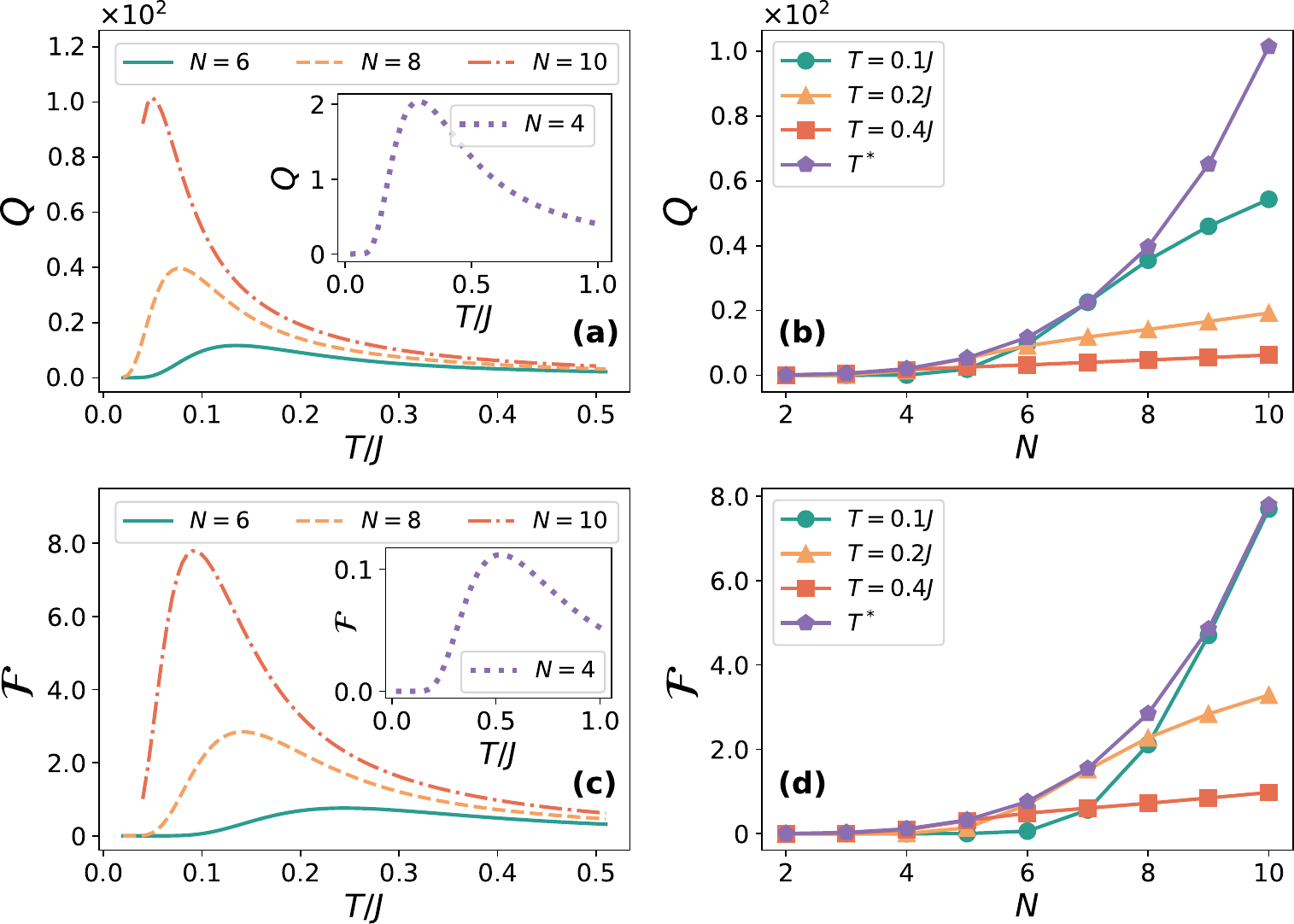} 
\caption{(a) Quantum Fisher information $\mathcal{Q}$ as a function of temperature $T$ for several system sizes $N$. (b) Quantum Fisher information $\mathcal{Q}$ as a function of the system size $N$ for several temperatures $T$. (c) Classical Fisher information $\mathcal{F}$ as a function of temperature $T$ for various system sizes $N$. (d) Classical Fisher information $\mathcal{F}$ as a function of system size $N$ for different temperatures $T$.}\label{fig_qfi_cfi} 
\end{figure}

In Fig.~\ref{fig_qfi_cfi}(a), the quantum Fisher information of the Gibbs state in Eq.~\eqref{eq_gibbs_state} is plotted as a function of the temperature $T$ for various system sizes $N$. As the figure shows, the quantum Fisher information increases with the system size $N$. This implies that the inherent nature of the many-body probe enhance thermometric capabilities. To understand the behavior of the quantum Fisher information for the Heisenberg spin-chain at thermal equilibrium, one can examine the average energy as the system size increases under two extreme temperature limits, namely: $T\rightarrow 0$ and $T\rightarrow\infty$. The former case involves the ground state of the system, resulting in an average energy of $\langle H \rangle = -J(N - 1)$, while the latter scenario yields an asymptotic value of $\langle H \rangle \rightarrow 0$. The transition between these two limits gives rise to an inflection point and, consequently, to the peak observed in the quantum Fisher information. As the average energy of the ground state decreases with the increasing system size $N$, the rate for which the transition between these two extreme cases occur becomes more pronounced. Consequently, the quantum Fisher information increases significantly, yet maintaining the same behavior. Note that in these two limits, $T\rightarrow 0$ and $T\rightarrow\infty$, the quantum Fisher information becomes effectively independent of the unknown parameter $T$ (i.e., $\mathcal{Q}\rightarrow 0$). However, the quantum Fisher information reaches its maximum at low temperature values. A temperature range highly relevant from a practical perspective.

In Fig.~\ref{fig_qfi_cfi}(b), the quantum Fisher information is presented as a function of the system size $N$ for various temperature values $T$. As evident from the figure, a fixed temperature $T$ leads to a consistent increase in the quantum Fisher information, implying that spin-chains with a larger number of particles indeed contribute to enhanced thermometry. Interestingly, the maximum quantum Fisher information with respect to temperature, denoted as $\mathcal{Q}_\mathrm{max}=\mathcal{Q}(T=T^*)$, for a given $N$, exhibits a clear non-linear behavior as the system size increases. This scenario is particularly valuable for local estimation theory, where significant prior information with respect to the unknown parameter is available.

To extract all the information content about the temperature, one needs to perform measurements in the energy basis of the entire spin-chain as defined in Eq.~\eqref{eq_gibbs_state}~\cite{PhysRevE.83.011109}. However, gaining access to and resolving the complete spectra of the Heisenberg spin-chain is not always feasible. Therefore, we evaluate the classical Fisher information using a sub-optimal yet accessible measurement basis:
\begin{equation}
    \Pi_{\pmb{\gamma}^{(N)}} = \bigotimes_{j=1}^{N} \frac{\mathbb{I}_j + (-1)^{\gamma_j} \sigma_z^j}{2}, \hspace{1 cm} \gamma_j = \text{0 or 1.} \label{eq_povm_cfi_total}
\end{equation}
Using this measurement basis, it becomes straightforward to calculate all the probabilities associated to measurement outcomes $\pmb{\gamma}^{(N)}=(\gamma_1,\gamma_2,\ldots,\gamma_N)$ from $(0_1,0_2,\ldots,0_N)$ to $(1_1,1_2,\ldots,1_N)$, and subsequently, determine the corresponding classical Fisher information.

In Fig.~\ref{fig_qfi_cfi}(c), the classical Fisher information is plotted as a function of temperature $T$ for various system sizes $N$. As the figure shows, the advantages of utilizing a Heisenberg many-body thermometry persist even with this sub-optimal measurement basis. In addition, the behavior of the classical Fisher information mirrors that observed for the quantum Fisher information, although with a reduced factor of approximately ${\sim}10$.

In Fig.~\ref{fig_qfi_cfi}(d), we present the classical Fisher information as a function of the system size $N$ for various temperatures $T$. Similar to the quantum Fisher information discussed earlier, we denote $\mathcal{F}_\mathrm{max}=\mathcal{F}(T=T^*)$ as the maximum classical Fisher information with respect to the temperature for a given $N$. As observed in the figure, with the sub-optimal measurement basis considered in this context, a distinct non-linear increase occurs with the increase of the system size.

As discussed above, the strong thermalization regime is constrained to the probe's heat capacity. In our case, the heat capacity of the quantum many-body probe increases with the system size, yet takes no advantage of the sequential measurements thermometry scheme as the probe in between measurements is strongly thermalized once again into a Gibss state. Thus, corresponding to an effective copy of the probe's initial state.

\subsection{Weak thermalization regime, $\kappa \ll J$}\label{subsec_seq_measu_thermo_A}

For this regime of thermalization, one can approximate the quantum many-body probe to evolve in the limit of vanishingly weak probe thermalization, namely when the Gibbs state evolves according to the unitary time operator $U(\tau)=e^{-i\tau H}$ (in Planck units, $\hbar = 1$) between consecutive measurements. Without loss of generality, we measure a local spin at site $N$ via
\begin{equation}
    \Pi_{\gamma_j}^{(N)} = \frac{\mathbb{I}_N + (-1)^{\gamma_j} \sigma_z^N}{2}, \hspace{1 cm} \gamma_j = \text{0 or 1,}
\end{equation}
undergoing fixed evolution time $J\tau=N$ between consecutive measurements.

\begin{figure}
\includegraphics[width=\linewidth]{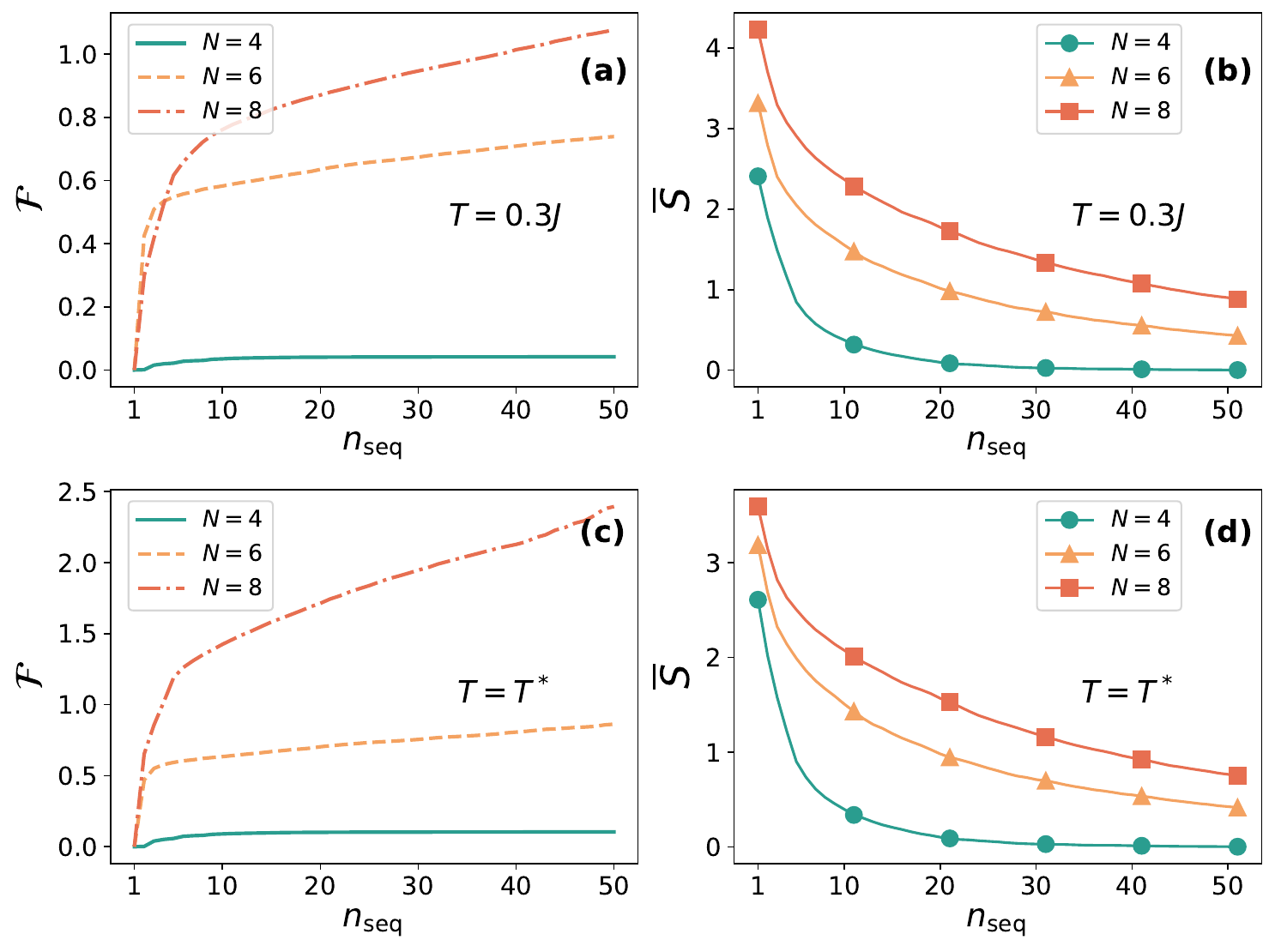} 
\caption{(a) and (c) show the classical Fisher information $\mathcal{F}$ as a function of $n_\mathrm{seq}$ for various system sizes $N$ and fixed $T=0.3J$ and $T=T^*$, respectively. (b) and (d) show the averaged von Neumann entropy $\overline{\mathcal{S}}$ as a function of $n_\mathrm{seq}$ for different $N$ and fixed $T=0.3J$ and $T=T^*$, respectively.}\label{fig2_unitary_CFI_vn_entropy} \end{figure}

In Fig.~\ref{fig2_unitary_CFI_vn_entropy}(a), we plot the classical Fisher information $\mathcal{F}$ as a function of the number of sequential measurements $n_\mathrm{seq}$ for various system sizes $N$ and a fixed temperature $T=0.3J$. As seen from the figure, a noticeable increase in the classical Fisher information is evident for a small number of sequences $n_\mathrm{seq}\sim 10$. Subsequently, the classical Fisher information grows significantly slower as the number of sequences increases, indicating that the information gain with respect to temperature becomes weaker. To understand this behavior, we calculate the von Neumann entropy of the entire system, averaged over $\mu$ total number of Monte Carlo trajectories, as follows:
\begin{equation}
    \overline{\mathcal{S}} = \frac{1}{\mu}\sum_{j=1}^\mu \mathcal{S}_j,
\end{equation}
where
\begin{equation}
    \mathcal{S}_j= - \mathrm{Tr}[\rho_j^{(n_\mathrm{seq})} \log \rho_j^{(n_\mathrm{seq})}]
\end{equation}
is the von Neumann entropy computed for the state $\rho_j^{(n_\mathrm{seq})}$, which is the density matrix for trajectory $j$ where $n_\mathrm{seq}$ measurements have been performed on the probe. In Fig.~\ref{fig2_unitary_CFI_vn_entropy}(b), we plot the averaged entropy $\overline{\mathcal{S}}$ as a function of the number of sequences $n_\mathrm{seq}$ for various system sizes $N$ and the given temperature $T=0.3J$. As observed in the figure, the averaged entropy decreases monotonically for all cases, starting from a highly mixed state and converging towards zero, indicating a pure state $\overline{\mathcal{S}}=0$. Hence, the system undergoes purification with each measurement performed on the probe. As the system size increases, the cost of purifying the system also rises.

Same behaviors can be found across other temperature values. In Fig.~\ref{fig2_unitary_CFI_vn_entropy}(c), we plot the classical Fisher information $\mathcal{F}$ as a function of the number of sequences $n_\mathrm{seq}$ for various system sizes $N$ as $T=T^*$. Here, $T^*$ is the temperature at which the quantum Fisher information of the thermalized probe in the absence of sequential measurements reaches its maximum. As the figure shows, the same non-linearity occurs as $n_\mathrm{seq}$ increases, in agreement with similar works on the matter~\cite{PhysRevResearch.5.043273}. In Fig.~\ref{fig2_unitary_CFI_vn_entropy}(d), we plot the averaged von Neumann entropy $\overline{\mathcal{S}}$ as a function of the number of sequential measurements $n_\mathrm{seq}$ for various system sizes $N$ at $T=T^*$. As evident from the figure, the action of measuring the probe consecutively causes the probe to purify. Thus, in the scenario of vanishingly weak probe thermalization, the classical Fisher information saturates towards a constant value as $n_\mathrm{seq}\gg 1$. This suggests that after the system undergoes purification, an additional sequential measurement on the probe does not contribute significantly to the information gain regarding the temperature.

\subsection{Intermediate thermalization regime, $\kappa \sim J$}\label{subsec_finite}
While previous thermalization regimes show benefits by considering a quantum many-body probe, the question on quantum-enhanced thermometry is constrained due to the heat capacity of the probe (strong thermalization regime) and the purification of the probe (weak thermalization regime). Intuitively, leaving the probe to compete between these two cases, might harness the sequential measurements thermometry more efficiently, giving rise to surpassing the thermometry achieved by energy measurements over the entire system. 

\begin{figure}[t]
\includegraphics[width=\linewidth]{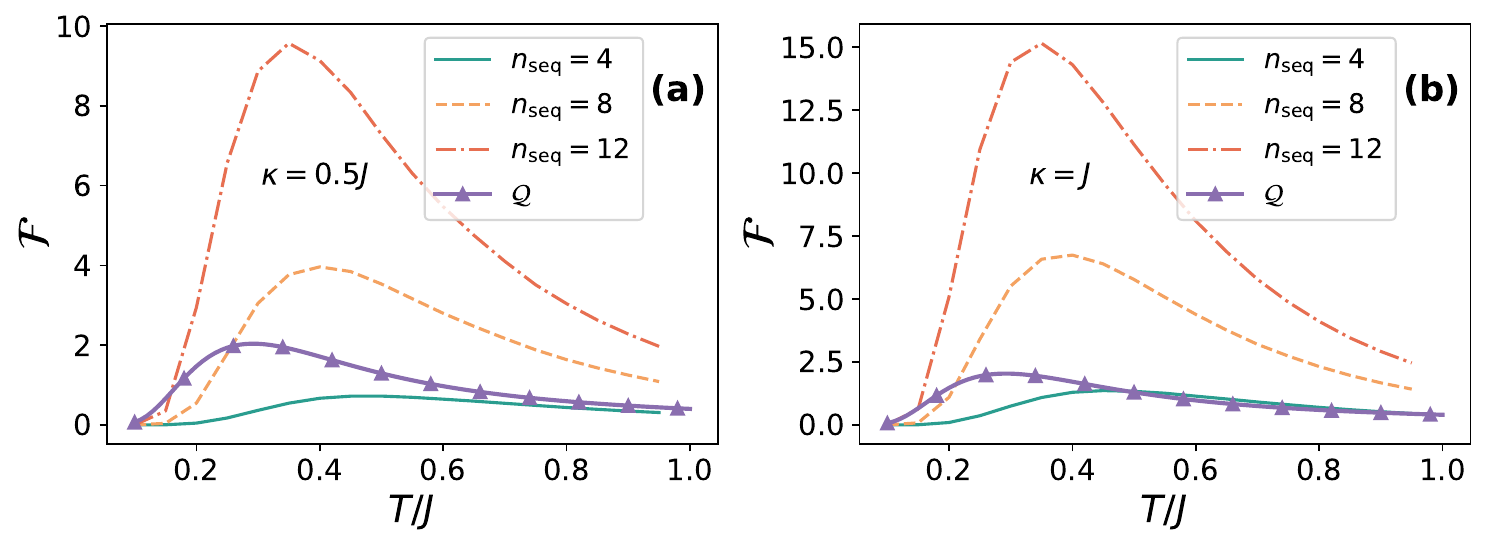} 
\caption{The classical Fisher information $\mathcal{F}$ is depicted as a function of temperature $T$ for various numbers of sequential measurements $n_\mathrm{seq}$. In (a) [(b)], we consider the thermalization strength as $\kappa=0.5J$ [$\kappa=J$]. Here, the system size is set to $N=4$, and the evolution time between consecutive measurements is fixed at $J\tau=N$. The quantum Fisher information (denoted as $\mathcal{Q}$) achieved with equilibrium probes is plotted for comparison.} \label{fig_F_vs_T_nn_and_full} 
\end{figure}

In Figs.~\ref{fig_F_vs_T_nn_and_full}(a)-(b), we plot the classical Fisher information as a function of temperature $T$ for several numbers of sequential measurements $n_\mathrm{seq}$ and two different thermalization strengths $\kappa=0.5J$ and $\kappa=J$, respectively. We consider $J\tau=N$ as the time in between consecutive measurements, with $N=4$. As seen from the figures, with the increase of the number of sequential measurements $n_{\mathrm{seq}}$, the resulting Fisher information increases across all the temperature points. Remarkably, for both thermalization strengths, the classical Fisher information obtained through the sequential measurements protocol highly exceeds the quantum Fisher information $\mathcal{Q}$ achieved with equilibrium probes with just a few ($n_\mathrm{seq}\sim 8$) sequential measurements. This highlights the remarkable power of sequential measurements in thermometry: with a few undemanding measurements on the local spin, one can achieve precision surpassing that of equilibrium probes, which require challenging energy measurements with full probe accessibility.

\begin{figure}[t]
\includegraphics[width=\linewidth]{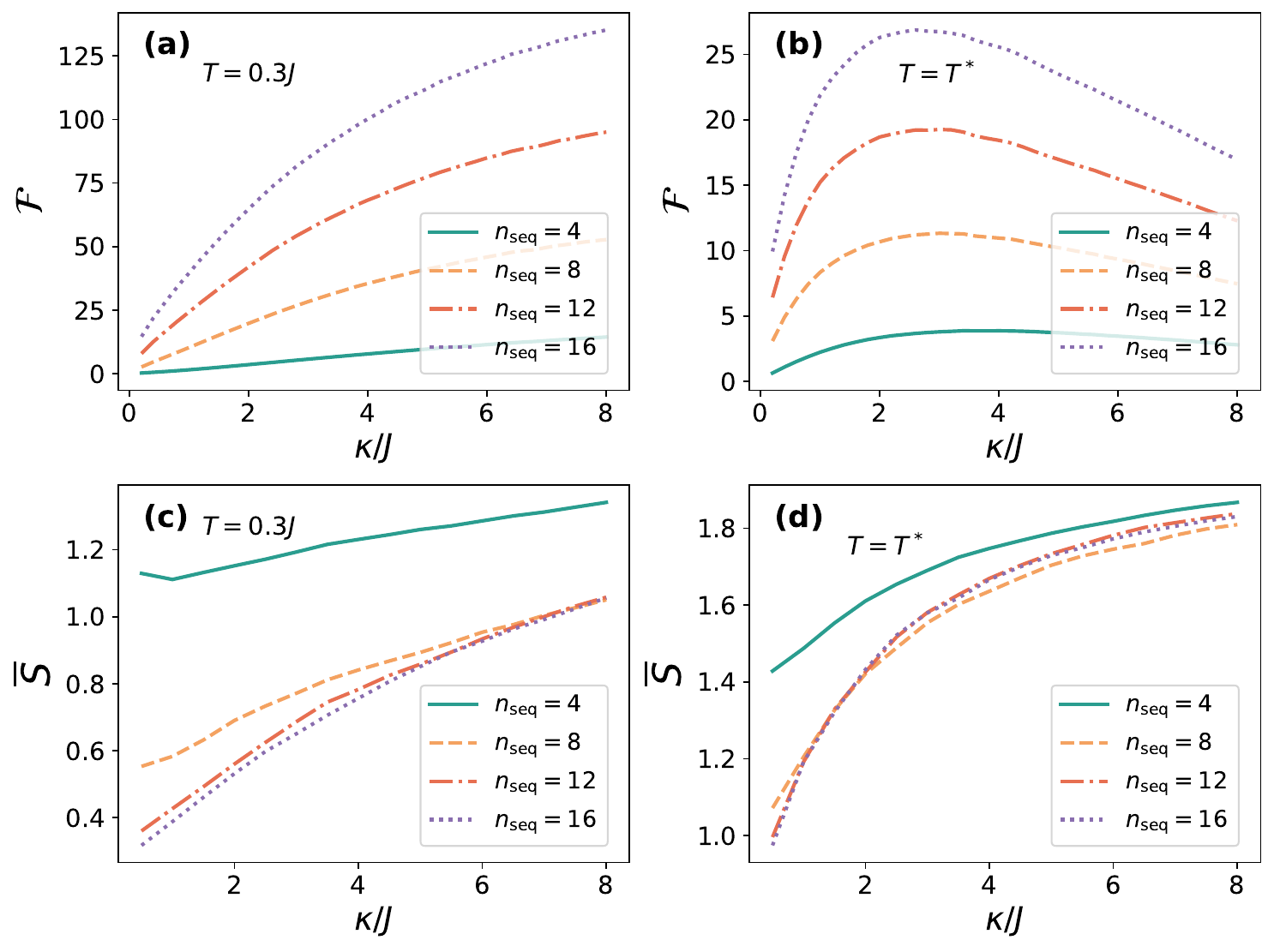} 
\caption{Simulation of the sequential measurement thermometry scheme considering a range of decay rates $\kappa$. (a) and (b) compare the classical Fisher information $\mathcal{F}$ as a function of the decay rate $\kappa$ for several number of sequential measurements $n_\mathrm{seq}$ for $T=0.3J$ and $T=T^*$, respectively. (c) and (d) compare the averaged von Neumann entropy of the entire system $\overline{\mathcal{S}}$ as a function of the decay rate $\kappa$ for several number of sequential measurements $n_\mathrm{seq}$ for $T=0.3J$ and $T=T^*$, respectively. Here, the system size is set to $N=4$, and the
evolution time between consecutive measurements is fixed at
$J\tau=2N$.}\label{fig_F_vs_gamma_full}
\end{figure}

Further analysis of the intermediate thermalization regime is found in Fig.~\ref{fig_F_vs_gamma_full}, where we investigate the classical Fisher information and the averaged von Neumann entropy for different thermalization strengths. To efficiently simulate the system, we set the system size to $N=4$. Furthermore, to more clearly observe the behavior of the curves, we adjust the evolution time interval between measurements to $J\tau=2N$. In Fig.~\ref{fig_F_vs_gamma_full}(a) and Fig.~\ref{fig_F_vs_gamma_full}(b), we illustrate the classical Fisher information as a function of the decay rate $\kappa$ for several numbers of sequential measurements $n_\mathrm{seq}$ for $T=0.3J$ and $T=T^*$, respectively. As seen from the figure, allowing the system to thermalize can achieve higher classical Fisher information with respect to the number of sequential measurements, thus harnessing sequential measurement thermometry. In addition, as evident from Fig.~\ref{fig_F_vs_gamma_full}(b), an \textit{optimal} $\kappa$ value is found to give the highest classical Fisher information. This situation is also expected for the case of $T=0.3J$ for a very large decay rate, namely $\kappa\gg J$ (not shown in the figure).

In Figs.~\ref{fig_F_vs_gamma_full}(c)-(d), we compare the averaged von Neumann entropy $\overline{\mathcal{S}}$ as a function of the decay rate $\kappa$ for various numbers of sequential measurements $n_\mathrm{seq}$ at temperatures $T=0.3J$ and $T=T^*$. As both figures show, the overall increasing behavior provides evidence of the probe's tendency to thermalize as $\kappa$ increases. To understand this behavior, we recall that the probe evolves between consecutive measurements. Therefore, this peak arises as a competition between the system evolving unitarily ($J \gg \kappa$) and the system being fully thermalized between consecutive measurements ($J \ll \kappa$). In the former case, the probe purifies at a faster rate, leading to the loss of temperature encoding in the quantum state. In the latter case, the probe thermalizes more rapidly, becoming an effective copy of the original Gibbs state. An intermediate scenario gives rise to the optimal situation in which the probe, in between consecutive measurements, carries the maximum information content regarding the temperature, reaching the highest value of the classical Fisher information. The fact that the probe becomes strongly thermalized as $\kappa$ increases also explains the universal overlap of the averaged von Neumann entropy $\overline{\mathcal{S}}$ as $\kappa$ increases, regardless of the number of sequential measurements performed on the probe.

\begin{figure}[t]
\includegraphics[width=\linewidth]{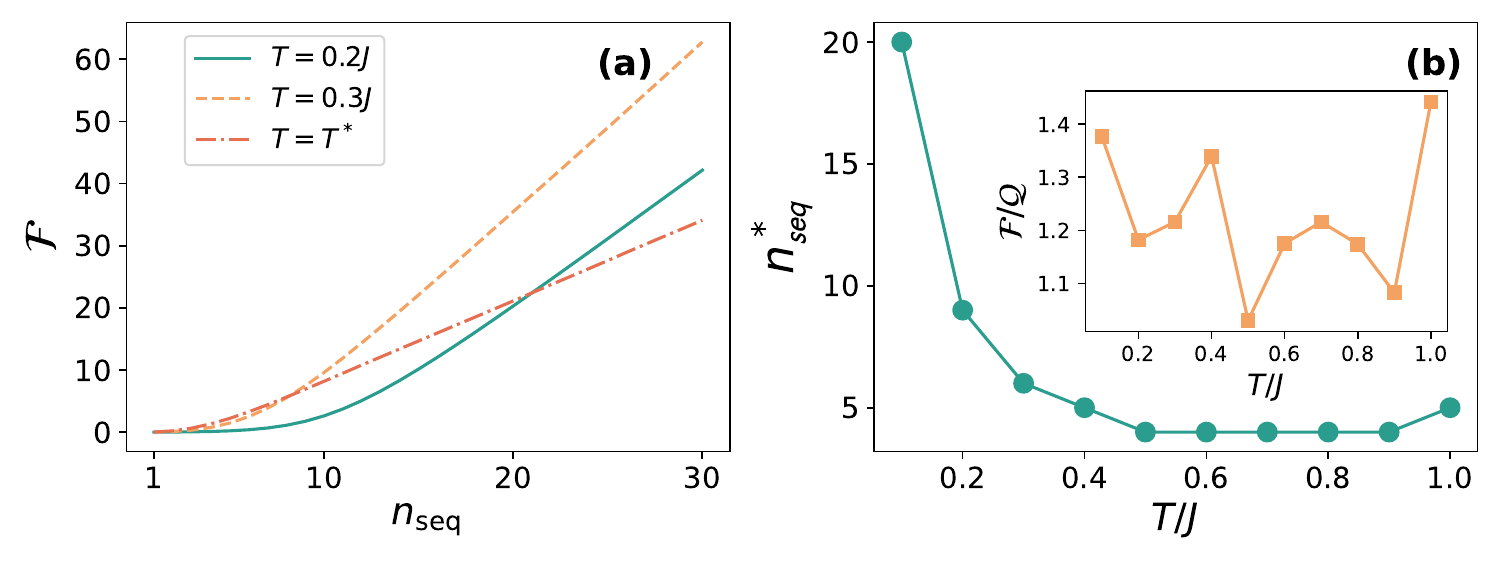} 
\caption{(a) Classical Fisher information $\mathcal{F}$ as a function of $n_{\mathrm{seq}}$ for different temperatures. (b) The number of sequential measurements $n_{\mathrm{seq}}^*$ needed to exceed the quantum Fisher information obtained by probes at equilibrium as a function of temperature $T$. The inset in (b) depicts the ratio $\mathcal{F}/\mathcal{Q}$ between the classical Fisher information $\mathcal{F}$ at the number of measurements $n_{\mathrm{seq}}^*$ over the quantum Fisher information $\mathcal{Q}$. The thermalization rate is set to $\kappa=J$, the system size is set to $N=4$, and the evolution time between consecutive measurements is fixed at $J\tau=N$.}\label{figure_extra_full_jumps}
\end{figure}

In Fig.~\ref{figure_extra_full_jumps}(a), we plot the classical Fisher information $\mathcal{F}$ as a function of the number of sequential measurements $n_\mathrm{seq}$ at several temperatures. As shown in the figure, a clear nonlinear behavior for a short number of sequential measurements $n_\mathrm{seq}$ is observed, while a linear trend with respect to $n_\mathrm{seq}$ is exhibited as $n_\mathrm{seq}$ increases. Most notably, in Fig.~\ref{figure_extra_full_jumps}(b) we quantify the minimum number of sequential measurements $n_\mathrm{seq}^*$ needed to surpass the optimal thermometry precision achieved in energy measurements with full accessibility. As seen from the figure, for the temperature range $0.1J \leq T \leq 1.0J$, the needed number of sequential measurements $n_\mathrm{seq}^*$ increases (decreases) as the temperature decreases (increases). This can be understood as in the limit towards $T\rightarrow 0$, the Gibbs state will be an eigenstate of the system, namely the spins parallel to each other (the triplet state). 

To quantify the advantage of a probe subjected to sequential measurement thermometry over a thermalized probe with energy measurements, we consider the ratio $\mathcal{F}/\mathcal{Q}$. Here, $\mathcal{F}$ represents the classical Fisher information obtained with the minimum number of sequential measurements $n_\mathrm{seq}^*$ required to exceed the quantum Fisher information $\mathcal{Q}$ obtained from the equilibrium probe. In the inset of Fig.~\ref{figure_extra_full_jumps}(b), by performing $n_\mathrm{seq}^*$ number of sequential measurements, one could surpass the quantum Fisher information by $\sim 20\%$, for most of the temperatures shown in the inset of Fig.~\ref{figure_extra_full_jumps}(b). 

Note that our scheme is minimal, i.e., no optimization on the scheme is performed. Improvement upon optimizing our scheme using quantum control and feedback control can improve the minimal $n_\mathrm{seq}$ to surpass optimal thermometry in energy basis. Indeed, the surpassing of the conventional optimal thermometry bound via local undemanding measurements using sequential measurements thermometry is the main result of our work.

\section{Conclusions}\label{sec_conc}

The estimation of temperature has primarily been explored at equilibrium and within the conventional quantum parameter estimation scheme, where a probe is reset after each measurement. In this work, we introduce two main differences: (i) we employ a non-conventional sequential measurements scheme in which the probe will be different after each measurement due to wave function collapse, and (ii) we exploit non-equilibrium dynamics to enhance accuracy. We investigate a many-body probe with Heisenberg interaction in three distinct thermalization scenarios: strong, weak, and intermediate thermalization regimes. Our results reveal that the competition between quantum state purification due to consecutive measurements and thermalization due to interaction with the reservoir leads to higher thermometry precision as the number of sequential measurements increases. Notably, our results demonstrate that it is possible to surpass conventional optimal thermometry, achievable through energy measurement over the entire system at thermal equilibrium, by using a sequence of local single-qubit measurements on the probe. This constitutes the main result of our work. Furthermore, our sequential measurements thermometry operates under minimal control constraints. Hence, by fine-tuning parameters such as the evolution time interval, it is feasible to further enhance the protocol's sensitivity.

\section*{acknowledgements}

A.B. acknowledges support from the National Natural Science Foundation of China (Grants No. 12050410253, No. 92065115, and No. 12274059), and the Ministry of Science and Technology of China (Grant No. QNJ2021167001L). V.M. thanks the National Natural Science Foundation of China (Grants No. 12050410251 and No. 12374482).

\appendix
\counterwithin{figure}{section}
\section{Temperature Estimation}\label{sec_t_est}

The classical Fisher information determines the sensitivity of an unknown parameter, such as temperature, for a specific POVM. Thus, it represents the bound achievable for a particular measurement basis. However, this bound does not, in actuality, provide the uncertainty of an estimated value. To complete the estimation procedure, it is necessary to input the collected measurement data into an estimator. This maps the collected data into the parameter space~\cite{paris2009quantum}, providing the actual uncertainty of the estimated value. To demonstrate the actual temperature uncertainty of our sequential measurement thermometry protocol, we use a Bayesian estimator that takes the (correlated) collected data from each quantum trajectory of length $n_{\mathrm{seq}}$ and provides an estimated temperature. The Bayes' rule is formulated as follows:
\begin{equation}
P(T|\pmb{\Gamma})=\frac{P(\pmb{\Gamma}|T)P(T)}{P(\pmb{\Gamma})},\label{eq_bayes-theorem-sm}
\end{equation}
where $P(T|\pmb{\Gamma})$, known as \textit{posterior}, is the conditional probability for the temperature $T$ given a set of measurement outcomes $\pmb{\Gamma}$; $P(T)$ represents the \textit{prior} information about the unknown temperature $T$. For the sake of simplicity, we assume the prior to be uniformly distributed; $P(\pmb{\Gamma}|T)$, known as \textit{likelihood}, is the conditional probability for the measurement outcomes $\pmb{\Gamma}$ assuming the unknown parameter $T$; and finally, the denominator $P(\pmb{\Gamma})$ is a normalization factor such that the posterior is a valid probability distribution, namely $\int_{T'} P(T'|\pmb{\Gamma})dT'= 1$.

In sequential measurements sensing, the observed data of $M$ trajectories is a collection of quantum trajectories given by:
\begin{equation}
\pmb{\Gamma}{=}\{\pmb{\gamma}_1,\pmb{\gamma}_2,\cdots,\pmb{\gamma}_M\},
\end{equation}
where $\pmb{\Gamma}$ is an array of vector components $\pmb{\gamma}_k$ (representing the $k$th quantum trajectory), each with $n_\mathrm{seq}$ spin measurement outcomes, and $M$ is the number of times the probe has been reset. The likelihood function then considers the probable occurrence of each quantum trajectory in the $M$ resetting runs, as follows~\cite{PhysRevLett.129.120503}:
\begin{equation}
P(\bm{\Gamma}|T)=\frac{M!}{k_1!k_2!\cdots k_{2^{n_\mathrm{seq}}}!}\prod_{j=1}^{2^{n_\mathrm{seq}}}\left[p(\pmb{\gamma}_j|T)\right]^{k_j},
\end{equation}
where $p(\pmb{\gamma}_j|T)$ is the conditional probability for quantum trajectory $\pmb{\gamma}_j$ assuming $T$, $k_1,{\cdots},k_{2^{n_\mathrm{seq}}}$ represent the number of times that the sequence $\pmb{\gamma}_1=(0_1,0_2,\ldots,0_{n_\mathrm{seq}})$ to $\pmb{\gamma}_{2^{n_\mathrm{seq}}}=(1_1,1_2,\ldots,1_{n_\mathrm{seq}})$ occurs in the entire sampling data set $M$. Here, $0_k$ and $1_k$ are the $\sigma_z$ eigenvalues for the $k$th measurement instance. It is worth emphasizing that $p(\pmb{\gamma}_j|T)$ requires classically simulating the probability distributions for all possible sequences from $\pmb{\gamma}_1$ to $\pmb{\gamma}_{2^{n_\mathrm{seq}}}$ over a relevant range of $T$. This might entail a significant computational cost as $n_\mathrm{seq}$ increases. We now present the two representative thermometry scenarios addressed in our work, namely when the probe-bath evolves within a vanishingly weak thermalization strength and when the probe-bath thermalizes with a finite strength.

\begin{figure}[t]
\includegraphics[width=\linewidth]{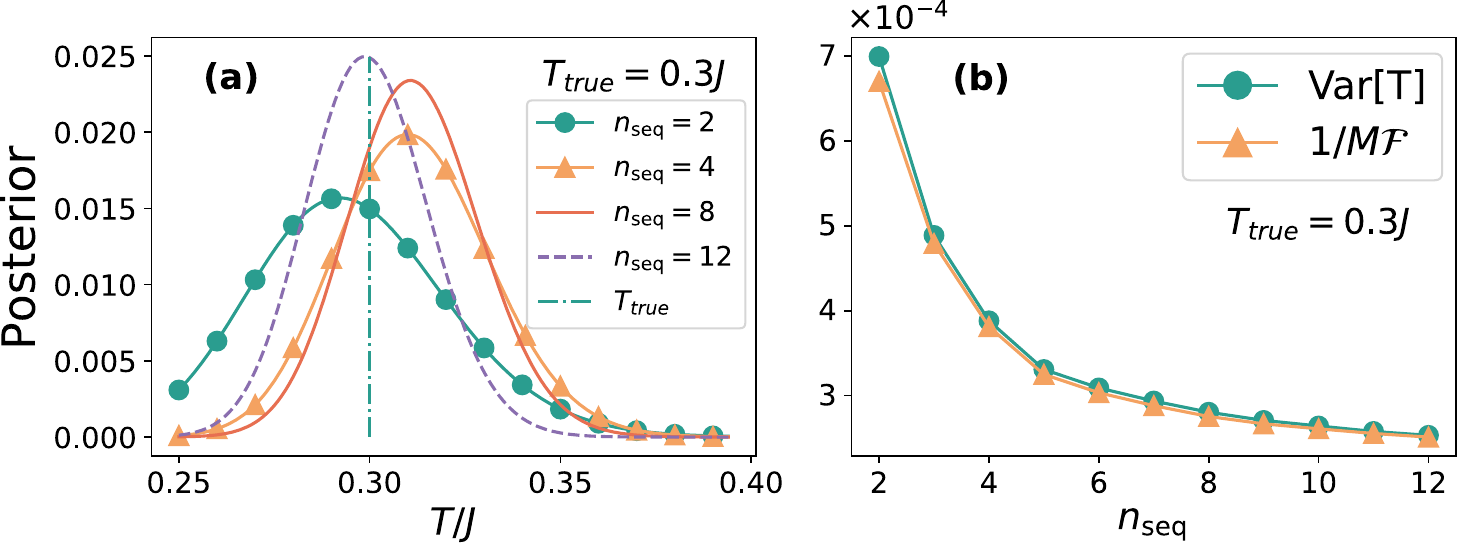} 
\caption{Unitary dynamics limit: (a) Posterior as a function of temperature $T$ for different number of sequential measurements $n_\mathrm{seq}$. (b) Comparison between the variance of the temperature estimation $\mathrm{Var}[T]$ and the inverse of the classical Fisher information $\mathcal{F}^{-1}$ as a function of $n_\mathrm{seq}$. We aim to estimate a true (unknown) temperature of $T=0.3J$ using a sampling of $M=5000$ quantum trajectories. The system size is set to $N=8$.} \label{fig3_bayesian} 
\end{figure}

\begin{figure}[t]
\includegraphics[width=\linewidth]{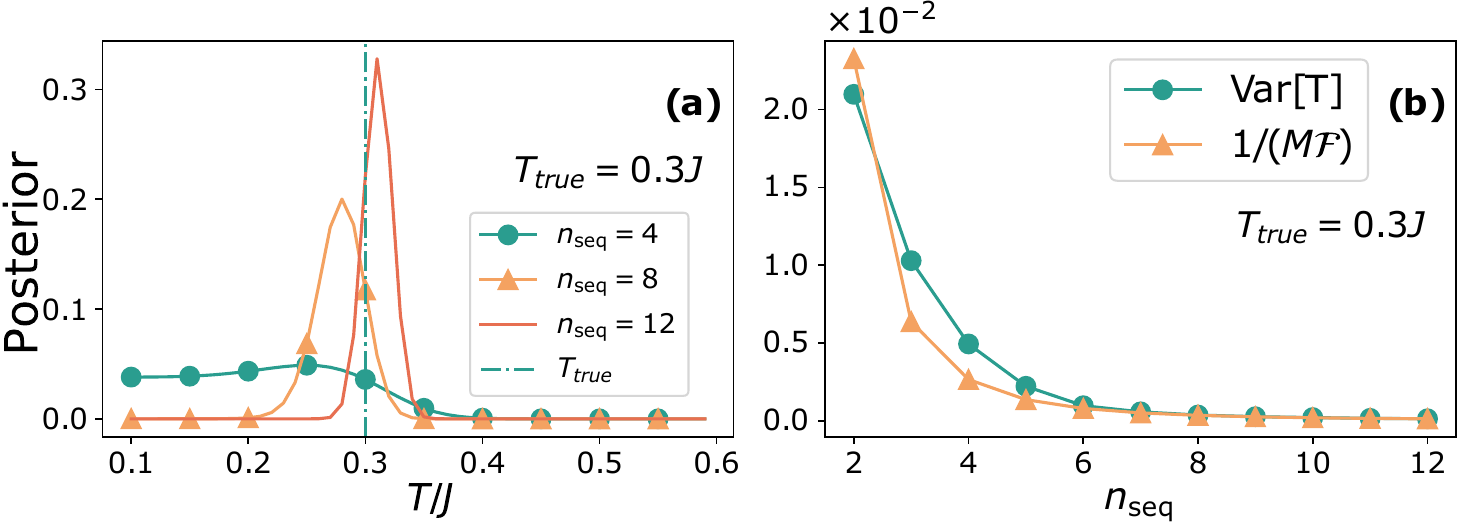} 
\caption{Open dynamics in the presence of finite probe-bath thermalization strength: (a) Posterior as a function of temperature $T$ for different numbers of sequential measurements $n_\mathrm{seq}$. (b) Comparison between the variance of the temperature estimation $\mathrm{Var}[T]$ and the inverse of the classical Fisher information $\mathcal{F}^{-1}$ as a function of $n_\mathrm{seq}$. We aim to estimate a true (unknown) temperature of $T=0.3J$ using a sample size of $M=500$ quantum trajectories. Note that in this scenario, the posterior is narrower than in the unitary dynamics case [see Fig.~\ref{fig3_bayesian}(a)], remarkably achieved using a smaller number of quantum trajectory samples. Other parameters are $N=4$ and $\kappa(\omega)=\kappa=J$.}\label{fig_NN_posterior} 
\end{figure}

Let us start the analysis for the vanishingly weak thermalization strength, i.e. the unitary dynamics limit. In Fig.~\ref{fig3_bayesian}(a), we show the posterior as a function of the temperature $T$ for several number of sequential measurements $n_\mathrm{seq}$. Here, we assume that we want to estimate the unknown (true) temperature of $T=0.3J$ utilizing $M=5000$ quantum trajectories (i.e., the probe has been reset $M=5000$ times). As the figure indicates, the posterior function tends to center around the true value. Most notably, its spread (or variance of $T$, $\mathrm{Var}[T]$) becomes narrower, thus explicitly showing a reduction in temperature uncertainty as $n_\mathrm{seq}$ increases. In Fig.~\ref{fig3_bayesian}(b), we compare the variance of the temperature $\mathrm{Var}[T]$ and the inverse of the classical Fisher information $\mathcal{F}^{-1}$ as a function of the number of sequential measurements $n_\mathrm{seq}$. As seen from the figure, with the increasing of $n_\mathrm{seq}$, the uncertainty of temperature reduces significantly, with the saturation in agreement with the purification of the system due to the consecutive measurements performed on the probe [see the discussion in Sec.~\ref{subsec_seq_measu_thermo_A}]. Remarkably, with a finite number of samples $M=5000$, the variance and the inverse of the classical Fisher information almost overlaps, demonstrating both the saturation of the Cram\'{e}r-Rao inequality $M\mathrm{Var}[T] \approx \mathcal{F}^{-1}$ and the efficiency of the chosen Bayesian estimation in the presence of correlated measurement outcomes.

In the case of open dynamics with finite thermalization strength, we have demonstrated that the intermediate thermalization regime leads to enhanced thermometry capabilities---see Sec.~\ref{subsec_finite} for details. To further illustrate temperature estimation using Bayesian estimation method, in Fig.~\ref{fig_NN_posterior}(a), we show the posterior as a function of temperature $T$ for various numbers of sequential measurements $n_\mathrm{seq}$. Here, $T=0.3J$ represents the true unknown temperature, and we consider $M=500$ quantum trajectories. As the figure illustrates, two important observations can be made from the posterior function: (i) the posterior peaks at higher values, resulting in a significant reduction in the uncertainty of the temperature; and (ii) this reduction in uncertainty can be achieved with a lower number of quantum trajectories, specifically $M=500$, which is one order of magnitude less compared to the unitary dynamics case with $M=5000$. To explicitly observe such a reduction in temperature uncertainty, in Fig.~\ref{fig_NN_posterior}(b), we compare the variance of the temperature estimation $\mathrm{Var}[T]$ with the inverse of the classical Fisher information $\mathcal{F}^{-1}$ as functions of $n_\mathrm{seq}$. As seen from the figure, one achieves $M\mathrm{Var}[T] \approx \mathcal{F}^{-1}$ with significantly fewer required quantum trajectories. This observation indicates that allowing the system to thermalize in the intermediate thermalization regime gives rise to an enhanced estimation of the temperature.

\bibliography{Thermometry_seq}

\end{document}